\documentclass{article}
\usepackage[dvips]{graphicx}
\usepackage{amssymb}
\usepackage{amsbsy}
\usepackage{amsthm}
\usepackage{enumerate}

\theoremstyle{plain}
\newtheorem{thm}{Theorem}[section]
\newtheorem{lem}[thm]{Lemma}
\newtheorem{cor}[thm]{Corollary}

\theoremstyle{definition}
\newtheorem{defn}[thm]{Definition}
\theoremstyle{remark}

\newtheorem*{prf}{Proof}

\title{Several Classes of Concatenated Quantum Codes: Constructions and Bounds\footnote{
This paper was presented in part at the IEICE Technical Meeting on Information Theory,
Nagoya, Japan, March 2006.}}
\author{
Hachiro FUJITA%
  \thanks{
Superrobust Computation Project,
University of Tokyo,
5-1-5 Kashiwanoha, Kashiwa-shi, Chiba 277-8561, Japan.
E-mail: {fujita@it.k.u-tokyo.ac.jp}
  }
}

\date{}

\begin{document}
\maketitle

\begin{abstract}
In this paper we present several classes of asymptotically good concatenated quantum codes
and derive lower bounds on the minimum distance and rate of the codes.
We compare these bounds with the best-known bound of Ashikhmin--Litsyn--Tsfasman and Matsumoto.
We also give a polynomial-time decoding algorithm for the codes that can decode up to
one fourth of the lower bound on the minimum distance of the codes.
\end{abstract}

\section{Introduction}
Quantum error correction is a basic technique for transmitting quantum information reliably
over a noisy quantum channel.
Many explicit constructions of quantum error-correcting codes have been proposed so far.
Some of the best-known code constructions are
the CSS code construction of Calderbank and Shor~\cite{CS} and Steane~\cite{Ste}
and the stabilizer code construction of Gottesman~\cite{Got96,Got97} and
Calderbank~{\it et al.}~\cite{CRSS97,CRSS98}.
CSS codes are constructed by using classical error-correcting codes and
have a simple decoding algorithm.
On the other hand, stabilizer codes are the most general class of quantum error-correcting codes known to date
and can be understood by using a theory of additive codes over GF(4), the Galois field with four elements.

As in classical coding theory, we want to construct quantum codes with large minimum distance.
More generally, we want to construct asymptotically good quantum codes that have minimum distance proportional to
the code length.
Ashikhmin~{\it et al.}~\cite{ALT} and Chen~{\it et al.}~\cite{CLX01} constructed asymptotically good quantum codes
based on algebraic geometry codes.
Later, Matsumoto~\cite{Mat} improved the bound of Ashikhmin~~{\it et al.}~\cite{ALT}.

In classical coding theory, code concatenation~\cite{For66a}
is a basic method for constructing good error-correcting codes
and most of the known asymptotically good binary codes are constructed by code concatenation~\cite{Dum}.
In 1971, Zyablov~\cite{Zya} constructed a family of asymptotically good binary codes
by concatenating Reed--Solomon (RS) outer codes with good binary inner codes,
and obtained the bound on the minimum distance of the codes, which is called
the Zyablov bound.

In the quantum setting, code concatenation is also effectively used
to construct good quantum error-correcting codes,
although concatenation is mainly used for fault-tolerant quantum computation~\cite{KL}.
Gottesman states code concatenation in his PhD thesis and gives the stabilizer of a quantum code constructed
by concatenating the five-qubit code with itself.
Calderbank~{\it et al.}~\cite{CRSS98} also remark concatenated codes and Rains~\cite{Rai} proves the so-called
product bound of concatenated codes.

In this paper we present several classes of concatenated quantum codes, more specifically quantum analogues
of the Zyablov codes, generalized concatenated codes, and the Blokh--Zyablov codes,
and give the bounds on the minimum distance of these codes.
We also give a quantum analogue of the Katsman--Tsfasman--Vl\u{a}du\c{t} bound based on algebraic geometry codes.

This paper is organized as follows:
In Section 2, we review stabilizer codes and the concept of code concatenation,
give a quantum analogue of the Zyablov codes,
which is constructed by concatenating quantum Reed--Solomon outer codes~\cite{GGB}
with good stabilizer inner codes,
and derive a lower bound on the minimum distance of the codes.
In Section 3, we extend the quantum Zyablov codes to the quantum version of generalized concatenated codes
and improve the quantum Zyablov bound. Furthermore, we give a quantum analogue of the Blokh--Zyablov bound.
In Section 4, we present a class of concatenated quantum codes based on algebraic geometry codes and
give a quantum analogue of the Katsman--Tsfasman--Vl\u{a}du\c{t} bound.
In Section 5, we discuss the decoding of the concatenated quantum codes constructed in this paper.
Based on the result of Hamada~\cite{Ham02},
we can show that the quantum Zyablov codes achieve the capacity attainable by
general stabilizer codes in time polynomial in block length.
This coding scheme should be contrasted with the random stabilizer coding scheme
which requires exponential time complexity to achieve the same capacity, although the error exponent of
the general stabilizer codes is much better that that of the quantum Zyablov codes.
In Section 6, we give the conclusion of the paper.

\section{Code concatenation and the quantum Zyablov bound}
We denote the finite field (Galois field) with $q$ elements by $\mathbb{F}_{q}$ (not by GF($q$)),
where $q$ is a prime power, and the $q$-ary entropy function by
\[
H_{q}(x)=-x\log_{q}\frac{x}{q-1}-(1-x)\log_{q}(1-x), \quad 0<x<1.
\]
If $q=2$, then $H_{2}(x)$ is the binary entropy function and denoted by $H(x)$ for simplicity.
Following the line of Calderbank~{\it et al.}~\cite{CRSS98}, we explain stabilizer quantum codes
(quantum codes for short)
and the construction of the concatenated quantum codes.
We also give the quantum Gilbert--Varshamov bound for the general stabilizer quantum codes.
Stabilizer quantum codes can be related with self-orthogonal additive codes over $\mathbb{F}_{4}$.
Let $\omega$ be a primitive element of $\mathbb{F}_{4}$ that satisfies $\omega^2=\omega+1$.
Then $\mathbb{F}_{4}=\{0,1,\omega,\omega^2\}$.
We define conjugation by $\bar{x}:=x^2$ for $x\in \mathbb{F}_{4}$.
Let $\boldsymbol{u}=(u_{1},u_{2},\cdots,u_{n})$,
$\boldsymbol{v}=(v_{1},v_{2},\cdots,v_{n})\in \mathbb{F}_{4}^{n}$.
We define the trace inner product of $\boldsymbol{u}$ and $\boldsymbol{v}$ as:
\[
\langle \boldsymbol{u},\boldsymbol{v}\rangle:=\sum_{i=1}^{n}(u_i\overline{v_i}+\overline{u_i}v_i).
\]
A classical {\it additive} code over $\mathbb{F}_{4}$ of length $n$ is
an additive subgroup of $\mathbb{F}_{4}^{n}$.
If $C$ is an $(n,2^{n-k})$ additive code, its {\it trace-dual} (simply {\it dual}) of $C$ is defined to be
\[
C^{\perp}:=\{\boldsymbol{u}\in \mathbb{F}_{4}^{n} \mid \textrm{$\langle \boldsymbol{u},\boldsymbol{v}\rangle=0$
for all $\boldsymbol{v}\in C$} \}.
\]
Then $C^{\perp}$ is an $(n,2^{n+k})$ additive code. If $C \subseteq C^{\perp}$, then $C$ is said to be
{\it self-orthogonal}.
For $\boldsymbol{u}\in \mathbb{F}_{4}^{n}$, we define the {\it weight} of $\boldsymbol{u}$ to be the number of
nonzero components of $\boldsymbol{u}$.
Let $C$ be an $(n,2^{n-k})$ self-orthogonal additive code.
Then the codes $C \subseteq C^{\perp}$
correspond to a quantum code $Q$ that encodes $k$ qubits in $n$ qubits.
If there are no vectors of weight $<d$ in $C^{\perp}\setminus C$, then $Q$ can correct
up to $\lfloor \frac{d-1}{2}\rfloor$ errors~\cite[Theorem 2]{CRSS98}
and $d$ is called the {\it minimum distance} of $Q$.
We denote by $[[n,k,d]]$ the parameters of such a quantum code $Q$.
Let $R_{Q}:=k/n$ and $\delta_{Q}:=d/n$.
\begin{thm}[\cite{EM,CRSS97}]~\label{Thm1}
For all sufficiently large $n$, there exists an $[[n,k,d]]$ quantum code satisfying
\begin{equation}
R_{Q}\ge 1-2H_{4}(\delta_{Q})=1-\delta_{Q}\log_{2}3-H(\delta_{Q}).\label{QGV}
\end{equation}
\end{thm}
Eq.~(\ref{QGV}) is called the {\it quantum Gilbert--Varshamov} (GV) bound,
since this bound is a quantum analogue of the GV bound
for classical binary (not necessarily linear) codes.
For self-containedness, we give a proof of Theorem~\ref{Thm1} in Appendix~\ref{AppendixA}.
The proof of Theorem~\ref{Thm1} is not constructive and it requires exponential time complexity
to find a quantum code satisfying Eq.~(\ref{QGV}).
Later, we compare this nonconstructive bound with our constructive ones.
We are now ready to introduce concatenated quantum codes~\cite{Got97,CRSS98}.
\begin{thm}[\cite{CRSS98}]~\label{Thm2}
If $Q_1$ is an $[[n_1m,k]]$ quantum code such that the associated $(nm,2^{nm+k})$ code has minimum
nonzero weight $d_1$ considered as a block code over an alphabet of size $4^{m}$,
and $Q_2$ is an $[[n_2,m,d_2]]$ quantum code, then encoding each block of $Q_1$ using $Q_2$
produces an $[[n=n_1n_2,k,d\ge d_1d_2]]$ concatenated quantum code.
\end{thm}
The proof of the above theorem will be clear from the construction of quantum Zyablov codes below.
A clear explanation of concatenated quantum codes can be found in~\cite[Sect.~IV]{Ham05}.
To construct quantum Zyablov codes, we need quantum Reed--Solomon codes introduced
by Grassl~{\it et al.}~\cite{GGB}.
Let $m$ be a positive integer.
A classical {\it Reed--Solomon} (RS) code $C_{\rm RS}$ of length $n=2^{m}-1$ over $\mathbb{F}_{2^{m}}$
is a cyclic code with generator polynomial
\[
g(x)=\prod_{i=0}^{d-2}\left(x-\alpha^{i}\right),
\]
where $\alpha$ is a primitive element of $\mathbb{F}_{2^{m}}$ and $2\le d\le 2^{m}-1$.
$C_{\rm RS}$ has dimension $k=n-d+1$ and minimum distance $d$.
RS codes are nonbinary codes.
We need a binary expansion of $C_{\rm RS}$.
\begin{defn}
Let $C$ be a linear code of length $n$ over $\mathbb{F}_{2^{m}}$, and
let $\mathcal{B}=\{b_1,\ldots,b_m\}$ be a basis of $\mathbb{F}_{2^{m}}$ over $\mathbb{F}_{2}$.
Then the {\it binary expansion} of $C$ with respect to the basis $\mathcal{B}$,
denoted by $\mathcal{B}(C)$, is the binary linear code of length $nm$ given by
\[
\mathcal{B}(C):=\left\{ (c_{ij})_{i,j}\in \mathbb{F}_{2}^{nm}\ \Big |\ 
\boldsymbol{c}=\left(\sum_{j=1}^{m}c_{ij}b_{j}\right)_{i}\in C \right\}.
\]
\end{defn}
For $k\le 2^{m-1}-1$, the RS code $C_{\rm RS}$ is self-orthogonal with respect to the standard
inner product of $\mathbb{F}_{2^{m}}$~\cite[Lemma 2]{GGB} and
so is the binary expansion $\mathcal{B}(C_{\rm RS})$ of $C_{\rm RS}$
with respect to a self-dual basis $\mathcal{B}$~\cite[Corollary 1]{GGB}.
Using the binary expansion of the RS code $C_{\rm RS}$ over $\mathbb{F}_{2^{m}}$
with parameters $[n,k,d]$, where $k\le 2^{m-1}-1$, with respect to a self-dual basis $\mathcal{B}$,
one can construct a $[[mn,m(n-2k)]]$ stabilizer quantum code $Q_{\rm RS}$ with associated additive codes
$C=\omega \mathcal{B}(C_{\rm RS})+\bar{\omega}\mathcal{B}(C_{\rm RS})$ and
$C^{\perp}=\omega \mathcal{B}(C_{\rm RS}^{\perp})+\bar{\omega}\mathcal{B}(C_{\rm RS}^{\perp})$
with parameters $(nm,2^{nm-m(n-2k)})$ and $(nm,2^{nm+m(n-2k)})$, respectively (see \cite[Theorem 9]{CRSS98}).
We call $Q_{\rm RS}$ a {\it quantum Reed--Solomon} (RS) code.
$Q_{\rm RS}$ has minimum distance at least $k+1$.
Although we describe quantum RS codes in terms of additive codes over $\mathbb{F}_{4}$,
quantum RS codes are a class of CSS codes~\cite{GGB}.

We now give the detail of the construction of concatenated quantum codes based on quantum RS codes.
Let $Q_1=Q_{\rm RS}$ be an $[[nm,m(n-2k)]]$ quantum RS code with associated codes
$C_1$, $C_{1}^{\perp}$ with parameters $(nm,2^{nm-m(n-2k)})$, $(nm,2^{nm+m(n-2k)})$ as above,
where $k\le 2^{m-1}-1$.
Then the associated code $C_{1}^{\perp}$ has minimum
nonzero weight $k+1$ considered as a block code over an alphabet of size $4^{m}$.
Let $Q_2$ be an $[[n_2,m,\delta_{2}n_2]]$ quantum code with associated additive codes
$C_2$, $C_2^{\perp}$ with parameters $(n_2,2^{n_2-m})$, $(n_2,2^{n_2+m})$
and suppose that $Q_2$ meets the quantum GV bound~(\ref{QGV}):
\begin{equation}
\delta_2=H_{4}^{-1}\left(\frac{1-r}{2}\right),\label{delta_2}
\end{equation}
where $r=m/n_2$.
Since $C_2^{\perp}/C_2$ has a natural symplectic structure,
there exists an inner-product-preserving map $\rho$ from $\mathbb{F}_{4}^{m}$ to $C_2^{\perp}/C_2$,
i.e., each $\boldsymbol{v}\in \mathbb{F}_{4}^{m}$ in 1-1 corresponds to
$\rho(\boldsymbol{v})\in C_2^{\perp}/C_2$ (see Appendix~\ref{AppendixB}).
We also denote by $\rho(\boldsymbol{v})$ a representative of the coset $\rho(\boldsymbol{v})$.
We define additive codes $\rho(C_1)$, $\rho(C_{1}^{\perp})$ as
\newpage
\begin{eqnarray*}
\rho(C_1)&\!:=\!&\{(\rho(\boldsymbol{v}_1)+\boldsymbol{u}_1,\rho(\boldsymbol{v}_2)+\boldsymbol{u}_2,\ldots,
\rho(\boldsymbol{v}_n)+\boldsymbol{u}_n) \mid \\
&&\quad (\boldsymbol{v}_1,\boldsymbol{v}_2,\ldots,\boldsymbol{v}_n)\in C_1,
\boldsymbol{u}_i\in C_2, 1\le i\le n\},\\
\rho(C_{1}^{\perp})&\!:=\!&
\{(\rho(\boldsymbol{v}_1)+\boldsymbol{u}_1,\rho(\boldsymbol{v}_2)+\boldsymbol{u}_2,\ldots,
\rho(\boldsymbol{v}_n)+\boldsymbol{u}_n) \mid \\
&&\quad (\boldsymbol{v}_1,\boldsymbol{v}_2,\ldots,\boldsymbol{v}_n)\in C_{1}^{\perp},
\boldsymbol{u}_i\in C_2, 1\le i\le n\}.
\end{eqnarray*}
Then it is easy to see that $\rho(C_1)$ and $\rho(C_{1}^{\perp})$ have parameters
$(nn_2,2^{nn_2-m(n-2k)})$ and $(nn_2,2^{nn_2+m(n-2k)})$, respectively,
and that $\rho(C_1)^{\perp}=\rho(C_{1}^{\perp})$.
The resulting quantum code $Q$ with associated codes $\rho(C_1)$, $\rho(C_{1}^{\perp})$,
called a {\it quantum Zyablov} code, has rate $R=r(1-2r')$, where $r'=k/n$, $0<r'<1/2$.
\begin{lem}\label{Lem3}
Let $\delta$ be the relative minimum distance of $Q$.
Then $\delta \ge \delta_{2}r'=r'H_{4}^{-1}\left(\frac{1-r}{2}\right)$.
\end{lem}
\begin{prf}
Let $\boldsymbol{c}=(\rho(\boldsymbol{v}_1)+\boldsymbol{u}_1,\ldots,
\rho(\boldsymbol{v}_n)+\boldsymbol{u}_n)\in \rho(C_{1}^{\perp})\setminus \rho(C_1)$, where
$(\boldsymbol{v}_1,\ldots,\boldsymbol{v}_n)\in C_{1}^{\perp}$ and
$\boldsymbol{u}_i\in C_2, 1\le i\le n$.
If $\boldsymbol{v}_i=\boldsymbol{0}$ for all $i$, then
$\rho(\boldsymbol{v}_i)+\boldsymbol{u}_i\in C_2$ for all $i$ and hence
$\boldsymbol{c}\in C_2\oplus \cdots \oplus C_2\subseteq \rho(C_1)$, which is a contradiction.
Hence $(\boldsymbol{v}_1,\boldsymbol{v}_2,\ldots,\boldsymbol{v}_n)$ is a nonzero codeword of $C_{1}^{\perp}$
and has at least $k+1$ nonzero components.
For each nonzero component $\boldsymbol{v}_i$,
$\rho(\boldsymbol{v}_i)+\boldsymbol{u}_i\in C_{2}^{\perp}\setminus C_{2}$ has weight
at least $\delta_{2}n_{2}$ and hence $\boldsymbol{c}$ has weight at least $\delta_{2}n_{2}(k+1)$.
This shows that the minimum distance of $Q$ is at least $\delta_{2}n_{2}(k+1)$.
From Eq.~(\ref{delta_2}) the statement follows.\qed
\end{prf}
For any given $R$, $0<R<1$, we maximize the relative minimum distance $\delta$ of $Q$
under the condition $R=r(1-2r')$.
From the above lemma we have
\begin{equation}
\delta \ge \max_{R<r<1}\frac{1}{2}\left(1-\frac{R}{r}\right)H_{4}^{-1}\left(\frac{1-r}{2}\right).\label{QZ}
\end{equation}
The maximum value of the right-hand side of Eq.~(\ref{QZ}) is taken at
\[
R=\frac{r^2}{1+2\log_4{\left(1- H_{4}^{-1}\left(\frac{1-r}{2}\right)\right)}}
\]
and does not vanish for any $R$, $0<R<1$.
We summarize the result in the following theorem.
\begin{thm}~\label{Thm3}
For any $R$, $0<R<1$, we can construct a family of asymptotically good concatenated quantum codes of
rate $R$
and relative minimum distance $\delta$ that satisfy Eq.~(\ref{QZ}).
\end{thm}
Eq.~(\ref{QZ}) is a quantum analogue of the Zyablov bound for classical concatenated
codes~\cite[Corollary~4.6]{Dum}.
This is the reason why we call $Q$ a quantum Zyablov code.

\begin{figure}[tb]
 \begin{center}
  \includegraphics[scale=0.65]{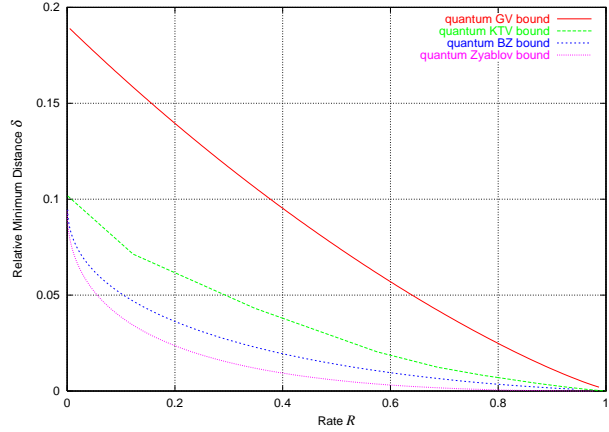}
  \caption{Comparison of the quantum GV, quantum KTV, quantum BZ and quantum Zyablov bounds.}
  \label{fig:1}
 \end{center}
\end{figure}

\section{Generalized concatenated quantum codes and the quantum Blokh--Zyablov bound}
In this section we present a class of generalized concatenated quantum codes,
which is a quantum analogue of classical generalized concatenated codes.
For the detail on classical generalized concatenated codes, see~\cite{Dum}.
We first give the construction of generalized concatenated quantum codes
and then derive minimum distance bounds for generalized concatenated quantum codes.
Let $s$ be a positive integer $\ge 2$.
To construct generalized concatenated quantum codes of order $s$, we need some notations.
Let $Q^{(i)}_{1}$, $1\le i\le s$,
be an $[[mn_1, m(n_1-2k_i)]]$ quantum RS code, where $m$ is a positive integer and $n_1=2^{m}-1$,
with associated codes $B_i$, $B_i^{\perp}$
with parameters $(mn_1, 2^{mn_1-m(n_1-2k_i)})$, $(mn_1, 2^{mn_1+m(n_1-2k_i)})$,
which are obtained from a binary expansion of a dual pair of classical RS codes.
Recall that the quantum RS code $Q^{(i)}_{1}$ has minimum distance at least $k_i+1$.
Furthermore, let $Q^{(j)}_{2}$, $1\le j\le s$,
be an $[[n_2, r_jn_2, \delta_jn_2]]$ quantum code, where $r_j=jm/n_2$, with associated codes $C_j$, $C_j^{\perp}$
with parameters $(n_2, 2^{n_2-jm})$, $(n_2, 2^{n_2+jm})$ such that
$C_{s}\subseteq C_{s-1}\subseteq \cdots \subseteq C_{1}$.
Consider the direct sums $B=B_1\oplus B_2 \cdots \oplus B_s$ and
$B^{\perp}=B_1^{\perp}\oplus B_2^{\perp} \cdots \oplus B_s^{\perp}$ (see~\cite{CRSS98}).
Note that the dual of $B$ is $B^{\perp}$ and that $B\subseteq B^{\perp}$.
We consider each $B_i^{\perp}$ as a block code over an alphabet of size $4^{m}$, as in the previous section, and
write a codeword $\boldsymbol{b}_i$ of $B_i^{\perp}$ as
$\boldsymbol{b}_i=(b_{i,1},b_{i,2},\cdots,b_{i,n_1})$, where $b_{i,l}\in \mathbb{F}_{4}^{m}$, $1\le l\le n_1$,
and we regard $\boldsymbol{b}=(\boldsymbol{b}_1,\boldsymbol{b}_2,\cdots,\boldsymbol{b}_{s})\in B^{\perp}$
as an $s\times n_1$ matrix over $\mathbb{F}_{4}^{m}$ whose $i$-th row is $\boldsymbol{b}_i$.

Let $\mathcal{B}_{j}$, $1\le j\le s$, be the set of $2m$ vectors of $C_{j}^{\perp}$ given
in Appendix~\ref{AppendixC}.
We now give the detail on the construction of a generalized concatenated quantum code of order $s$.
Consider the quotient map $\pi_{s}: C_s^{\perp}\to C_s^{\perp}/C_s$.
Since the quotient space $C_s^{\perp}/C_s$ that has a natural symplectic structure is isomorphic to
$\mathbb{F}_{4}^{sm}$ as a symplectic space,
there exists an inner-product-preserving map $\rho$ from $\mathbb{F}_{4}^{sm}$ to $C_s^{\perp}/C_s$.
We can assume that the $j$-th block of $\mathbb{F}_{4}^{sm}$ corresponds to span($\pi_{s}(\mathcal{B}_{j})$),
where $1\le j\le s$.
Using this $\rho$
we map the $j$-th column $\boldsymbol{c}_j$ of $\boldsymbol{b}\in B^{\perp}$ above, i.e.,
$\boldsymbol{c}_j=(b_{1,j},b_{2,j},\cdots,b_{s,j})\in \mathbb{F}_{4}^{sm}$, to
$\rho(\boldsymbol{c}_j)\in C_s^{\perp}/C_s$ and
obtain a map from $B^{\perp}$ to $\left(C_s^{\perp}/C_s\right)^{n_1}$, which is also denoted by $\rho$.
As in the previous section, from $B\subseteq B^{\perp}$ and $\rho$
we can construct additive codes $C\subseteq C^{\perp}$
with parameters $(n_1n_2, 2^{n_1n_2-\sum_{j=1}^{s}m(n_1-2k_j)})$,
$(n_1n_2, 2^{n_1n_2+\sum_{j=1}^{s}m(n_1-2k_j)})$.
The quantum code $Q$ with associated codes $C\subseteq C^{\perp}$ has rate $R$ given by
\begin{equation}
R=r-\frac{2r}{s}\sum_{j=1}^{s}r'_j,\label{GCQCR}
\end{equation}
where $r=r_s$ and $r'_j=k_j/n_1$.
\begin{lem}~\label{Lem2}
Suppose that $\delta_1\ge \delta_2\ge \cdots \ge \delta_s$.
Let $\delta$ be the relative minimum distance of $Q$. Then
\[
\delta\ge \min_{1\le j\le s}\delta_{j}r'_{j}.
\]
\end{lem}
\begin{prf}
Let $\boldsymbol{c}\in C^{\perp}\setminus C$.
As in Lemma~\ref{Lem3}, $\boldsymbol{c}$ is written as
$\boldsymbol{c}=\rho(\boldsymbol{b})+\boldsymbol{u}$, where
$\boldsymbol{b}=(\boldsymbol{b}_1,\boldsymbol{b}_2,\cdots,\boldsymbol{b}_{s})$ is a nonzero vector
of $B^{\perp}$ and $\boldsymbol{u}\in C_s^{n_1}$.
Suppose that $\boldsymbol{b}_{l}=\boldsymbol{0}$, $j+1\le l\le s$,
and $\boldsymbol{b}_{j}$ is the last nonzero row of $\boldsymbol{b}$.
Since $\boldsymbol{b}_{j}$ has at least $k_{j}+1$ nonzero components and
each encoded column of $\boldsymbol{b}$ is in $\pi_{s}(C_{j}^{\perp})$,
the weight of $\boldsymbol{c}=\rho(\boldsymbol{b})+\boldsymbol{u}$ is at least $\delta_{j}n_2(k_{j}+1)$.
Since $j$ ranges over the set $\{1,2,\ldots,s\}$, the minimum weight of $C^{\perp}\setminus C$
is at least $\min_{1\le j\le s}\delta_{j}n_2(k_{j}+1)$.
Hence the minimum distance of $Q$ is at least $\min_{1\le j\le s}\delta_{j}n_2(k_{j}+1)$
and the statement follows.\qed
\end{prf}
To derive a bound for asymptotically good
generalized concatenated quantum codes of order $s$,
we need a sequence of self-orthogonal additive codes $C_{i}$, $1\le i\le s$, over $\mathbb{F}_{4}$
of the same length satisfying the following two conditions:
\begin{enumerate}[i)]
\item
$C_{s}\subseteq C_{s-1}\subseteq \cdots \subseteq C_{1}$.
\item
Each quantum code $Q_{i}$ corresponding to the additive codes $C_{i}\subseteq C_{i}^{\perp}$
meets the quantum GV bound~(\ref{QGV}).
\end{enumerate}
Quantum codes $Q_{i}$ above have natural inclusion:
$Q_{1}\allowbreak\subseteq \allowbreak Q_{2}\allowbreak\subseteq \allowbreak\cdots \allowbreak\subseteq 
\allowbreak Q_{s}$.
The following lemma is a quantum version of ~\cite[Lemma~4.10]{Dum}.
\begin{lem}\label{Lem1}
Let $0<r_1<r_2<\cdots <r_s<1$.
For all sufficiently large $n$, there exist $s$ nested quantum codes $Q_{i}$, $1\le i\le s$,
with parameters $[[n,r_{i}n,\delta_{i}n]]$, which simultaneously meet the quantum GV bound
\begin{equation}
\delta_{i}\ge H_{4}^{-1}\left(\frac{1- r_{i}}{2}\right), \quad 1\le i\le s.\label{Lem1EQ}
\end{equation}
\end{lem}
The proof of Lemma~\ref{Lem1} is given in Appendix~\ref{AppendixD}.
The following is a quantum version of~\cite[Theorem 4.11]{Dum}:
\begin{thm}~\label{Thm4}
For any $r$, $0<r<1$, and $\delta<H_{4}^{-1}(\frac{1-r}{2})$
we can construct asymptotically good generalized
concatenated quantum codes of order $s$, relative minimum distance
at least $\delta/2$ and rate $R$ given by
\begin{equation}
R=r-\frac{r}{s}\sum_{j=1}^{s}\delta/H_{4}^{-1}\left(\frac{1}{2}\left(1-\frac{rj}{s}\right)\right).\label{GCQCR2}
\end{equation}
\end{thm}
\begin{prf}
We use the notations used in  the construction of the generalized
concatenated quantum code $Q$ above.
Recall that $r_j=jm/n_2$, $1\le j\le s$.
We choose the rate $r_s$ of the $s$-th inner quantum code $Q_{2}^{(s)}$ satisfying $r_s=r$.
Hence $0<r_1<r_2<\cdots <r_s=r$.
From Lemma~\ref{Lem2} the relative minimum distance of $Q$ is at least $\min_{1\le j\le s}\delta_{j}r'_{j}$.
From Lemma~\ref{Lem1} we can take $\delta_{j}=H_{4}^{-1}(\frac{1-r_j}{2})$.
For this value of $\delta_j$ we set $r'_j=\frac{\delta}{2H_{4}^{-1}\left(\frac{1-r_j}{2}\right)}$.
Note that $0<r'_j<1/2$.
Hence the relative minimum distance of $Q$ is at least $\delta/2$
and Eq.~(\ref{GCQCR}) gives the rate $R$ of $Q$.\qed
\end{prf}
Maximizing $R$ with respect to $r$ in Theorem~\ref{Thm4}, we obtain the following:
\begin{cor}
For any $\delta$, $0<\delta <H_{4}^{-1}(1/2)$, there exist a generalized
concatenated quantum code of order $s$, relative minimum distance at least $\delta/2$ and rate $R$ given by
\begin{equation}
R=\max_{0<r<1-2H_4(\delta)}r-\frac{r}{s}\sum_{j=1}^{s}\delta/H_{4}^{-1}\left(\frac{1}{2}\left(1-\frac{rj}{s}\right)\right).\label{GCQCR3}
\end{equation}
\end{cor}
Taking $s\to \infty$ in Theorem~\ref{Thm4}, we obtain the following:
\begin{cor}
For any $\delta$, $0<\delta <H_{4}^{-1}(1/2)$ and sufficiently large $s$, there exist a generalized
concatenated quantum code of order $s$, relative minimum distance at least $\delta/2$ and rate
close to
\begin{equation}
R=1-2H_4(\delta)-\delta \int_{0}^{1-2H_4(\delta)}\frac{dx}{H_{4}^{-1}\left(\frac{1-x}{2}\right)}.\label{QBZ}
\end{equation}
\end{cor}
Eq.~(\ref{QBZ}) is a quantum analogue of the Blokh--Zyablov (BZ) bound~\cite[Corollary 4.13]{Dum}.
We compare the quantum GV bound~(\ref{QGV}), the quantum Zyablov bound~(\ref{QZ}) and
the quantum BZ bound~(\ref{QBZ}) in Fig.~\ref{fig:1}.

\section{The quantum Katsman--\allowbreak Tsfasman--\allowbreak Vl\u{a}du\c{t} bound}
\begin{figure}[tb]
 \begin{center}
  \includegraphics[scale=0.65]{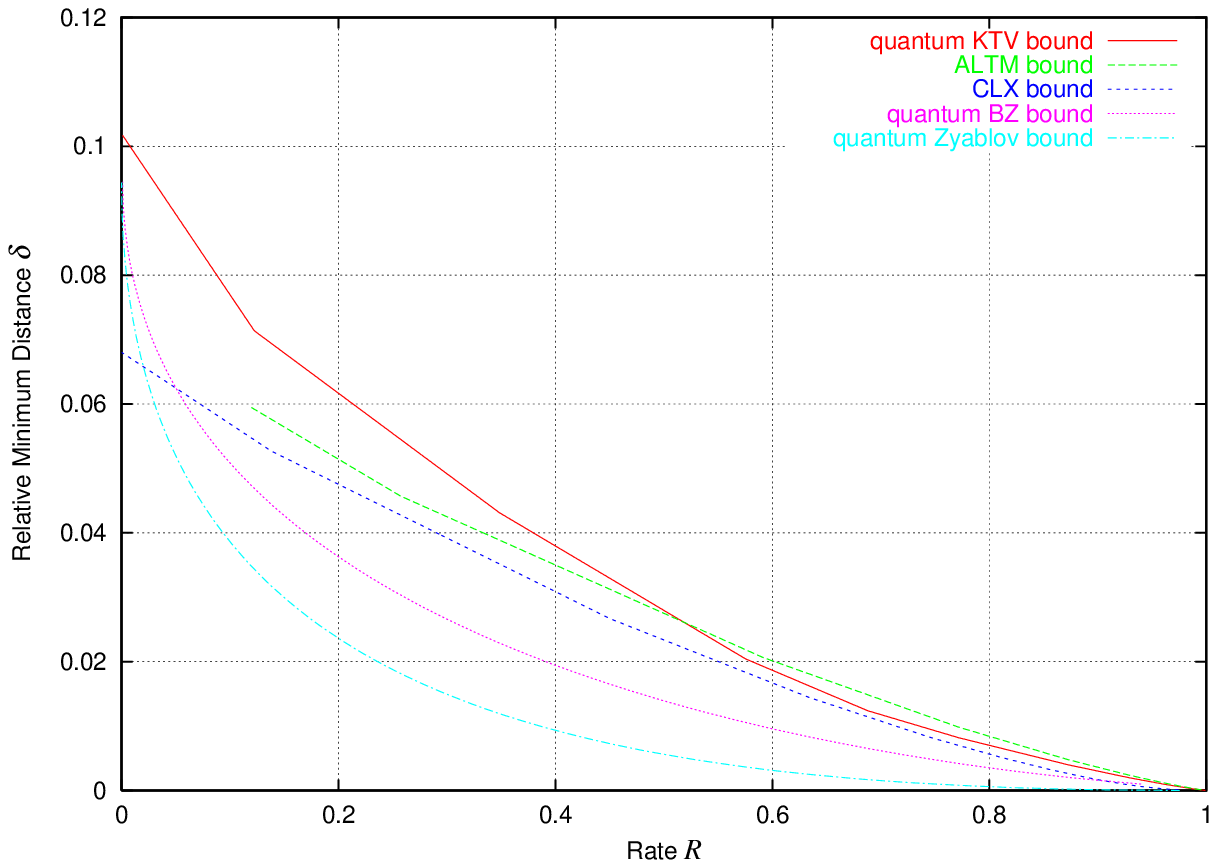}
  \caption{Comparison of the quantum KTV, ALTM, CLX, quantum BZ and quantum Zyablov bounds.}
  \label{fig:2}
 \end{center}
\end{figure}

In this section we present a class of concatenated quantum codes based on algebraic geometry codes.
We use the result of~\cite{Mat}.
Let $q=2^m$, where $m$ is a positive integer.
We need the Garcia--Stichtenoth tower of function fields over $\mathbb{F}_{q^2}$.
\begin{defn}[\cite{GS}]
Let $F_1:=\mathbb{F}_{q^2}(x_1)$ be the rational function field over $\mathbb{F}_{q^2}$.
For $i\ge 1$, we set
\[
F_{i+1}:=F_{i}(z_{i+1}),
\]
where $z_{i+1}$ satisfies the equation
\[
z_{i+1}^{q}+z_{i+1}=x_{i}^{q+1},
\]
with $x_{i}=z_{i}/x_{i-1}$, $i\ge 2$.
\end{defn}
Let $n_i=(q^{2}-1)q^{i-1}$.
The zero divisor of $x_1^{q^{2}-1}-1\in F_i$ consists of $n_i$ places of degree one
and hence we denote it by $P_{1}+P_{2}+\cdots +P_{n_i}$.
For a divisor $D$ of $F_i/\mathbb{F}_{q^2}$ with ${\rm supp}D\cap \{P_{1},P_{2},\cdots ,P_{n_i}\}=\emptyset$,
we define a linear code $C(D)$ over $\mathbb{F}_{q^2}$ as
\[
C(D)=\{(f(P_1),f(P_1),\cdots,f(P_{n_i}))\mid f\in \mathcal{L}(D)\}.
\]
Let $g_i$ be the genus of $F_i/\mathbb{F}_{q^2}$.
For each $i$ and $0\le j\le n_i/2-g_i$, there exists a divisor $H$ of $F_i/\mathbb{F}_{q^2}$ such that
the following two conditions hold:
\begin{enumerate}[i)]
\item
$C(H)\subseteq C(H)^{\perp}$.
\item
$C(H)^{\perp}$ has dimension $n_i/2+j$ and minimum distance at least $n_i/2-g_i+1-j$.
\end{enumerate}
For the explicit form of $H$ see \cite{Mat}.
As in the case of quantum RS codes, using the binary expansion of the codes $C(H)$, $C(H)^{\perp}$
over $\mathbb{F}_{q^2}$ we obtain additive codes $C$, $C^{\perp}$ with parameters
$(2mn_i,2^{2mn_i-4mj})$, $(2mn_i,2^{2mn_i+4mj})$.
The quantum code $Q_1$ with associated codes $C$, $C^{\perp}$ has parameters
$[[2mn_i,4mj,d_1\ge n_i/2-g_i+1-j]]$.
The rate $r_1$ of $Q_1$ is given by $r_1=2j/n_i$.
Let $Q_2$ be a quantum code with parameters $[[n,r_{2}n,\delta_{2}n]]$, where $r_2=2m/n$.
The concatenation of $Q_1$ with $Q_2$ gives an $[[nn_i, r_1r_2nn_i]]$ quantum code with relative minimum distance
$\delta$ that satisfies
\begin{equation}
\delta\ge \delta_2\left(\frac{1-r_1}{2}-\frac{g_i}{n_i}\right),\quad 0\le r_1\le 1-\frac{2g_i}{n_i}.\label{KTV1}
\end{equation}
Since $\lim_{i\to \infty}\frac{g_i}{n_i}=\frac{1}{q-1}$, taking $i\to \infty$ in~(\ref{KTV1})
leads to
\begin{equation}
\delta\ge \delta_2\left(\frac{1-r_1}{2}-\frac{1}{q-1}\right),\quad 0\le r_1\le \frac{q-3}{q-1}.\label{KTV2}
\end{equation}
Setting $R=r_1r_2$, we obtain
\begin{equation}
\delta\ge \delta_2\left(\frac{r_2-R}{2r_2}-\frac{1}{q-1}\right),\quad \frac{q-1}{q-3}R\le r_2\le 1.\label{KTV3}
\end{equation}
Eq.~(\ref{KTV3}) is a quantum analogue of the Katsman--Tsfasman--Vl\u{a}du\c{t} (KTV) bound \cite{KTV}.
Since there are many good quantum codes for short block lengths (see~\cite[Table III]{CRSS98}),
we can choose a good quantum code $Q_2$.
We optimized the right-hand side of Eq.~(\ref{KTV3}) with respect to $Q_2$ using the table in~\cite{CRSS98}
and obtained a quantum KTV bound, which is shown in Fig.~\ref{fig:1}.
In Fig.~\ref{fig:2} we compare several lower bounds for constructive quantum codes, that is,
the quantum KTV bound, the Ashikhmin--Litsyn--Tsfasman--Matsumoto (ALTM) bound~\cite{ALT,Mat},
the Chen--Ling--Xing (CLX) bound~\cite{CLX01},
the quantum BZ bound and the quantum Zyablov bound.
As can be seen from Fig.~\ref{fig:2},
the quantum KTV bound is superior to the ALTM bound for rates lower than about 0.5,
and the quantum BZ bound and the quantum Zyablov bound are superior to the CLX bound for very low rates.
The quantum KTV bound can be improved by using more efficient quantum codes not in the table in~\cite{CRSS98}.

\section{Decoding concatenated quantum codes}
In this section we give a decoding algorithm for concatenated quantum codes.
Let us now consider the quantum Zyablov code $Q$ constructed in Section~2, for example.
$Q$ is the concatenation of a quantum RS outer code $Q_1$ with an inner quantum code $Q_2$.
Suppose that $Q_1$ has associated codes $C_1$, $C_{1}^{\perp}$, and that
$Q_2$ has associated codes $C_2$, $C_{2}^{\perp}$.
The decoding algorithm consists of the following two steps:
\begin{enumerate}
\item {\it Inner Decoding:} For each inner code $Q_2$ of $Q$:
\begin{enumerate}
\item
Measure the generators of $C_2$ and
estimate the most likely errors from the measurement result.
\item
Correct the errors and decode the encoded data.
\end{enumerate}
\item {\it Outer Decoding:}
\begin{enumerate}
\item
Correct the remaining errors
in the quantum RS outer code $Q_1$ of $Q$ by using the CSS code structure of $Q_1$.
\item
Re-encode each block of $Q_1$ using $Q_2$, if necessary.
\end{enumerate}
\end{enumerate}
As in the case of classical concatenated codes~\cite[Theorem 5.1]{Dum},
it can be proven that the above decoding algorithm can correct up to $\delta N/4$ errors,
where $\delta$ is the lower bound on the relative minimum distance of $Q$ and
$N$ is the overall block length of $Q$, i.e., the number of qubits of $Q$.
We remark that the estimation of the most likely errors in $Q_2$ and the computation of the positions and
types (bit flip, phase flip, or both) of the errors in $Q_2$ can be done on a classical computer.
The estimation using exhaustive search
takes time exponential in the block length of the inner code, which is $\log N$.
Hence for each inner code the estimation complexity is $O(N)$ and the total complexity of estimating the
errors in all inner codes is $O(N^2)$.
The measurement and correction of all inner codes require $O(N(\log N)^2)$ quantum operations.
On the other hand, $Q_1$ can be decoded in $O(N^2)$ time using the Berlekamp--Massey algorithm
on a classical computer.
The syndrome computation and correction of $Q_1$ require $O(N^2)$ quantum operations.
Since any classical polynomial time algorithm can be done on a quantum computer in polynomial time,
the decoding algorithm above can be implemented on a quantum computer in polynomial time.
The above decoding algorithm applies also for concatenated quantum codes based on algebraic geometry codes.
As in the case of classical concatenated codes, it is possible to correct up to $\delta N/2$ errors
with generalized minimum distance decoding~\cite{For66b,Dum}.

Finally, we remark the fidelity of the quantum Zyablov code $Q$ above.
It is shown in~\cite{Ham02} that there exists a sequence of stabilizer quantum codes
of rate smaller than some quantity such that
the fidelity of a code in the sequence converges to $1$ exponentially as the block length grows.
Using this result, we can show that
if the block length of $Q$ is enough large, then the fidelity of $Q$ is arbitrarily close to $1$.
The proof is essentially the same as the classical counterpart~\cite[Theorem 4.15]{Dum}.

\section{Conclusion}
In this paper we have presented several constructions of asymptotically good concatenated quantum codes.
Concatenated quantum codes have simple structure and can be decoded efficiently in polynomial time.
Although we focus on the binary concatenated quantum codes, the extension to nonbinary concatenated quantum codes
is straightforward.

\section*{Acknowledgments}
The author would like to thank Prof.~H.~Yamamoto of the University of Tokyo for support.

\newpage
\appendix
\section*{Appendix}
\theoremstyle{plain}
\section{Proof of Theorem~\ref{Thm1}}~\label{AppendixA}
Following the argument in~\cite[Sect. V]{CS}, we prove Theorem~\ref{Thm1}.
Counting arguments used in the proofs below can also be found in \cite[Sect.~9.5]{HP}
and \cite[Sect.~7 in Chap.~17, Sect.~6 in Chap.~19]{MS}.
\begin{lem}\label{lemA1}
Let $\boldsymbol{v}$ be a nonzero vector of $\mathbb{F}_{4}^{n}$ and
$C$ self-orthogonal additive code over $\mathbb{F}_{4}$ of
length $n$ and dimension $k+1$ containing $\boldsymbol{v}$.
Then $C$ contains $2^{k}-1$  self-orthogonal additive subcodes of
dimension $k$ containing $\boldsymbol{v}$.
\end{lem}
\begin{prf}
We first remark that a subcode of $C$ is obviously self-orthogonal.
Let $C_{0}:=\{\boldsymbol{0},\boldsymbol{v}\}$. Then $C_{0}$ is a one dimensional additive subcode.
Consider the quotient map $C\to C/C_{0}$. Note that $C/C_{0}$ is a $k$-dimensional binary vector space.
Let $S$ be the set of $k$-dimensional subcodes of $C$
containing $\boldsymbol{v}$.
Then $S$ is identified with the set of $(k-1)$-dimensional subspaces of $C/C_{0}$.
The total number of $(k-1)$-dimensional subspaces of $C/C_{0}$ is $2^{k}-1$,
which completes the proof.\qed
\end{prf}

\begin{lem}\label{lemA2}
Let $\boldsymbol{v}$ be a nonzero vector of $\mathbb{F}_{4}^{n}$, and for $1\le k\le n$,
let $\sigma_k$ be the number of self-orthogonal additive codes of
length $n$ over $\mathbb{F}_{4}$ and dimension $k$ containing $\boldsymbol{v}$. Then
\[
\sigma_k=\prod_{i=1}^{k-1}\frac{2^{2(n-i)}-1}{2^{i}-1}
\]
\end{lem}
\begin{prf}
It is obvious that $\sigma_1=1$,
since a self-orthogonal additive code of length $n$ over $\mathbb{F}_{4}$ and dimension $1$
containing $\boldsymbol{v}$ is only $C_{0}:=\{\boldsymbol{0},\boldsymbol{v}\}$.
Let $C$ be a self-orthogonal additive code over $\mathbb{F}_{4}$ of
length $n$ and dimension $k$ containing $\boldsymbol{v}$, and
let $S$ be the set of self-orthogonal additive code over $\mathbb{F}_{4}$ of
length $n$ and dimension $k+1$ containing $C$.
If $C'\in S$, then $C\subseteq C'\subseteq C'^{\perp}\subseteq C^{\perp}$.
Consider the quotient map $C^{\perp} \to C^{\perp}/C$.
Then $S$ is identified with the set of $2^{2(n-k)}-1$ cosets of $C$ in $C^{\perp}$,
i.e., all the cosets other than $C$.
Let $C'\in S$.
By Lemma~\ref{lemA1}, $C'$ contains $2^{k}-1$ self-orthogonal additive subcodes
of dimension $k$ containing $\boldsymbol{v}$.
Therefore we have
\[
\sigma_{k+1}=\frac{2^{2(n-k)}-1}{2^{k}-1}\sigma_{k}.
\]
Using this recursion we obtain the expression for $\sigma_{k}$ as in the statement.\qed
\end{prf}


\begin{lem}\label{lemA3}
Let $\boldsymbol{v}$ be a nonzero vector of $\mathbb{F}_{4}^{n}$, and for $1\le k\le n-1$,
let $\tau_{k}$ be the number of self-orthogonal additive codes $C$ over $\mathbb{F}_{4}$
of length $n$ and dimension $k$ satisfying $\boldsymbol{v}\in C^{\perp}\setminus C$.
Then
\[
\tau_{k}=2^{k}\prod_{i=1}^{k}\frac{2^{2(n-i)}-1}{2^{i}-1}
\]
\end{lem}
\begin{prf}
Let $S$ be the set of self-orthogonal additive codes $C$ over $\mathbb{F}_{4}$
of length $n$ and dimension $k$ satisfying $\boldsymbol{v}\in C^{\perp}\setminus C$.
Then $\tau_{k}$ is the cardinality of the set $S$.
Pick $C\in S$ and consider the code $C'$ generated by $C$ and $\boldsymbol{v}$.
Then $C'$ is an self-orthogonal additive code of dimension $k+1$ containing $\boldsymbol{v}$.
$C'$ contains $2^{k+1}-1$ subcodes of dimension $k$, and from Lemma~\ref{lemA1},
$2^{k}-1$ of these subcodes contain $\boldsymbol{v}$.
Hence $C'$ contains $2^{k}$ subcodes of dimension $k$ not containing $\boldsymbol{v}$.
By Lemma~\ref{lemA2} the number of self-orthogonal additive code of dimension $k+1$
containing $\boldsymbol{v}$ is $\sigma_{k+1}$.
Hence we have $\tau_{k}=2^{k}\sigma_{k+1}$.\qed
\end{prf}

We are now ready to prove Theorem~\ref{Thm1}.
Let $\Phi$ be the set of all self-orthogonal additive codes over $\mathbb{F}_{4}$
of length $n$ and dimension $n-k$,
and let $\Phi^{\perp}:=\{C^{\perp} \mid C\in \Phi \}$.
From Lemmas~\ref{lemA2} and \ref{lemA3}, each nonzero vector $\boldsymbol{v}\in \mathbb{F}_{4}^{n}$
belongs to the same number $N$ of codes in $\Phi^{\perp}$, where $N=\sigma_{n-k}+\tau_{n-k}$.
Hence we have
\[
N\left(2^{2n}-1\right)=|\Phi^{\perp}|\left(2^{n+k}-1\right). 
\]
If $N\sum_{i=1}^{d-1}3^{i}{n \atopwithdelims() i}< |\Phi^{\perp}|$, i.e., 
\begin{equation}
\sum_{i=1}^{d-1}3^{i}{n \atopwithdelims() i}<\frac{2^{2n}-1}{2^{n+k}-1},
\end{equation}
then there exists an additive code $C^{\perp}\in \Phi^{\perp}$ that has minimum distance $\ge d$.

\section{The symplectic structure of $C_2^{\perp}/C_2$}\label{AppendixB}
Since $C_2^{\perp}\subseteq \mathbb{F}_{4}^{n_2}$ and $\mathbb{F}_{4}^{n_2}$ has the symplectic
inner product $\langle \cdot,\cdot \rangle$ defined in Section 2,
we define a symplectic inner product on $C_2^{\perp}/C_2$ as
\[
\langle \boldsymbol{u}+C_2, \boldsymbol{v}+C_2\rangle :=
\langle \boldsymbol{u}, \boldsymbol{v}\rangle,
\]
where $\boldsymbol{u}, \boldsymbol{v}\in C_{2}^{\perp}$.
It is easy to see that this definition does not depend on representatives $\boldsymbol{u}, \boldsymbol{v}$,
and that the induced form on $C_2^{\perp}/C_2$ is nondegenerate.

Let $\{b_1,b_2,\ldots,b_{m}\}$ be the standard basis of $\mathbb{F}_{4}^{m}$, i.e.,
$b_i=(\delta_{ij})_{j}$, where $\delta_{ij}$ is the Kronecker delta.
We set
\[
e_i=\omega b_i,\quad f_i=\bar{\omega}b_i,\quad 1\le i\le m.
\]
Then
\[
\langle e_i,f_j\rangle=\delta_{ij},\quad \langle e_i,e_j\rangle=0,\quad \langle f_i,f_j \rangle=0.
\]
It follows from the following lemma that
there exists an inner-product-preserving map from $\mathbb{F}_{4}^{m}$ to $C_2^{\perp}/C_2$.
\begin{lem}
For any $2m$-dimensional binary vector space $V$
with nondegenerate symplectic form
$(\cdot,\cdot)$,
there exists a basis $\{g_i, h_i, 1\le i\le m\}$ of $V$ over $\mathbb{F}_{2}$ such that
\[
(g_i,h_j)=\delta_{ij}, \quad (g_i,g_j)=0, \quad (h_i,h_j)=0.
\]
\end{lem}
\begin{prf}
We show the statement by induction on $m$.
In the case $m=1$,
pick a nonzero vector $v\in V$.
Since the form $(\cdot,\cdot)$ is nondegenerate, there exists another vector
$v'\in V$ that satisfies $(v,v')=1$.
$g_1=v$ and $h_1=v'$ give a desired basis.

Suppose that the statement holds for any $2(m-1)$-dimensional
binary vector space with a nondegenerate symplectic form.
Let $V$ be a $2m$-dimensional binary vector space
with nondegenerate symplectic form $(\cdot,\cdot)$.
As explained above, we can take vectors $g_1, h_1\in V$ that satisfies $(g_1,h_1)=1$.
Let $W={\rm span}\{g_1, h_1\}$ and consider the space $W^{\perp}$.
Since the form $(\cdot,\cdot)$ is nondegenerate, the dimension of $W^{\perp}$ is $2(m-1)$.
It is easy to see that $W\cap W^{\perp}=\{0\}$.
Hence $V$ is the direct sum of $W$ and $W^{\perp}$, and
the restriction of $(\cdot,\cdot)$ to $W^{\perp}$ gives a nondegenerate symplectic form on $W^{\perp}$.
By hypothesis
there exist a basis $\{g_i, h_i, 2\le i\le m\}$ of $W^{\perp}$ over $\mathbb{F}_{2}$ such that
\[
(g_i,h_j)=\delta_{ij}, \quad (g_i,g_j)=0, \quad (h_i,h_j)=0.
\]
Hence the set $\{g_i, h_i, 1\le i\le m\}$ gives a desired basis.\qed
\end{prf}

\section{The set $\mathcal{B}_{j}$}\label{AppendixC}
We first remark that
$C_{s}\subseteq C_{s-1}\subseteq \cdots C_{1}\subseteq C_{1}^{\perp}\subseteq C_{2}^{\perp}\subseteq
\cdots \subseteq C_{s}^{\perp}$.
Consider the quotient map $\pi_1:C_{1}^{\perp}\to C_{1}^{\perp}/C_{1}$.
We can take a basis $\{\bar{g}_{i}, \bar{h}_{i}, 1\le i\le m\}$ of $C_{1}^{\perp}/C_{1}$ 
over $\mathbb{F}_{2}$ such that
\begin{equation}
\langle \bar{g}_{i},\bar{h}_{j}\rangle=\delta_{ij},\quad \langle \bar{g}_{i},\bar{g}_{j}\rangle=0,\quad \langle \bar{h}_{i},\bar{h}_{j}\rangle=0.\label{EQ1}
\end{equation}
We choose $g_{i}, h_{i}\in C_{1}^{\perp}$, $1\le i\le m$, that satisfy
$\pi_1(g_{i})=\bar{g}_{i}$ and $\pi_1(h_{i})=\bar{h}_{i}$.
The following is easily checked:
\begin{equation}
\langle g_{i},h_{j}\rangle=\delta_{ij},\quad \langle g_{i},g_{j}\rangle=0,\quad \langle h_{i},h_{j}\rangle=0.
\label{EQ2}
\end{equation}
From Eq.~(\ref{EQ2}), it follows that
$\mathcal{B}_{1}=\{g_{i}, h_{i}, 1\le i\le m\}$ are linearly independent, and
that $C_{1}$ and $\mathcal{B}_{1}$ span $C_{1}^{\perp}$.

Next, consider the quotient map $\pi_2:C_{2}^{\perp}\to C_{2}^{\perp}/C_{2}$.
Let $\bar{g}_{i}=\pi_2(g_{i})$, $\bar{h}_{i}=\pi_2(h_{i})$, $1\le i\le m$.
Although we use the same notation above, $\bar{g}_{i}$ and $\bar{h}_{i}$ here are elements of
$C_{2}^{\perp}/C_{2}$.
Note that for $\bar{g}_{i}$, $\bar{h}_{i}$, $1\le i\le m$,
the same equations as in Eq.~(\ref{EQ1}) with a natural symplectic form on
$C_{2}^{\perp}/C_{2}$ are satisfied.
We can take $\bar{g}_{i}$, $\bar{h}_{i}$, $m+1\le i\le 2m$, of $C_{2}^{\perp}/C_{2}$ in such a way that
the same equations as in Eq.~(\ref{EQ1}) with $1\le i,j\le 2m$ are satisfied.
We choose $g_{i}, h_{i}\in C_{2}^{\perp}$, $m+1\le i\le 2m$, that satisfy
$\pi_2(g_{i})=\bar{g}_{i}$ and $\pi_2(h_{i})=\bar{h}_{i}$.
It is easily checked that the same equations as in Eq.~(\ref{EQ2}) with $1\le i,j\le 2m$ are satisfied.
As in the case of $C_1^{\perp}$, it is also easily checked that
$\mathcal{B}_{2}=\{g_{i}, h_{i}, 1\le i\le 2m\}$ are linearly independent, and
that $C_{2}$ and $\mathcal{B}_{2}$ span $C_{2}^{\perp}$.
We inductively define $\mathcal{B}_{i}$, $i\ge 3$, and obtain $\mathcal{B}_{s}=\{g_{i}, h_{i}, 1\le i\le sm\}$
that satisfy the same equations as in Eq.~(\ref{EQ2}) with $1\le i,j\le sm$.
Note that the vectors in $\mathcal{B}_{i}$ are linearly independent, and
that $C_{i}$ and $\mathcal{B}_{i}$ span $C_{i}^{\perp}$.

We redefine $\mathcal{B}_{j}$, $1\le j\le s$, as
$\mathcal{B}_{j}=\{g_{i}, h_{i}, (j-1)m+1\le i\le jm\}$
($\mathcal{B}_{1}$ is the same as above).

\section{Proof of Lemma~\ref{Lem1}}~\label{AppendixD}
We prove the case $s=2$ only.
The general case is a straightforward extension of the case $s=2$.
Let $0\le r_{1}\le r_{2}\le 1$ and sufficiently large $n$ be given.
Without loss of generality, we may assume that $k_{i}=r_{i}n$, $i=1,2$, are positive integers.
Let $C_{1}$ be a self-orthogonal additive code over $\mathbb{F}_{4}$
with parameters $(n, 2^{n-k_{1}})$ and suppose that its dual $C_{1}^{\perp}$
has minimum distance $d_1=\delta_{1}n$,
where $\delta_{1}=H_{4}^{-1}\left(\frac{1-r_{1}}{2}\right)$.
This is possible from Theorem~\ref{Thm1}.
We need to show that there exists a self-orthogonal additive code $C_{2}$ over $\mathbb{F}_{4}$
with parameters $(n, 2^{n-k_{2}})$ such that the following two conditions hold:
\begin{enumerate}[i)]
\item $C_{2}\subseteq C_{1}$.
\item The dual $C_{2}^{\perp}$ of $C_{2}$ has minimum distance $d_2=\delta_{2}n$,
where $\delta_{2}=H_{4}^{-1}\left(\frac{1- r_{2}}{2}\right)$.
\end{enumerate}
Consider the quotient map $\pi: \mathbb{F}_{4}^{n}\to \mathbb{F}_{4}^{n}/C_{1}^{\perp}$.
Note that the quotient space $\mathbb{F}_{4}^{n}/C_{1}^{\perp}$ is a $(n-k_{1})$-dimensional binary
vector space.
We define the weight of a coset to be the smallest weight of a vector in the coset, i.e.,
the weight of a coset is the weight of a coset leader.
Let $\Psi_{k_{2}-k_{1}}$ be the set of $(k_{2}-k_{1})$-dimensional subcodes of
$\mathbb{F}_{4}^{n}/C_{1}^{\perp}$.
If $C\in \Psi_{k_{2}-k_{1}}$, then $\widetilde{C}:=\pi^{-1}(C)$ is an $(n+k_{2})$-dimensional
additive code over $\mathbb{F}_{4}$ containing $C_{1}^{\perp}$.
Let $d:={\rm dist}(C)$. Then it is easy to see that ${\rm dist}(\widetilde{C})=\min \{d, d_1\}$.
(Consider the decomposition $\widetilde{C}=\cup_{\boldsymbol{v}\in \pi^{-1}(C)}\boldsymbol{v}+C_{1}^{\perp}$.)
Let $S$ be the set of cosets of nonzero weight smaller than $d$.
Since $S$ is in the image of the set of nonzero vectors of $\mathbb{F}_{4}^{n}$
of weight smaller than $d$ under $\pi$, we have
\[
|S|\le \sum_{i=1}^{d-1}3^{i}{n \atopwithdelims() i}.
\]
As in the proof of Theorem~\ref{Thm1}, we can prove that for $\boldsymbol{v}\not\in C_{1}^{\perp}$,
the number of codes in $\Psi_{k_{2}-k_{1}}$ containing $\boldsymbol{v}+C_{1}^{\perp}$
is independent of $\boldsymbol{v}+C_{1}^{\perp}$.
We denote the number by $N$.
Hence we have
\[
N(2^{n-k_{1}}-1)=|\Psi_{k_{2}-k_{1}}|(2^{k_{2}-k_{1}}-1).
\]
If $N|S|\le N\sum_{i=1}^{d-1}3^{i}{n \atopwithdelims() i}<|\Psi_{k_{2}-k_{1}}|$, i.e.,
\begin{equation}
\sum_{i=1}^{d-1}3^{i}{n \atopwithdelims() i}<\frac{2^{n-k_{1}}-1}{2^{k_{2}-k_{1}}-1},
\end{equation}
then there exists an additive code $C\in \Psi_{k_{2}-k_{1}}$ that has minimum distance $\ge d$.
A standard argument shows that we can take $d$ to be $d_2=\delta_{2}n$.
Since $\delta_{1}\ge \delta_{2}$, the corresponding $\widetilde{C}$ has minimum distance $d_2$.
We define $C_{2}:=\widetilde{C}^{\perp}$.
So $C_{2}^{\perp}=\widetilde{C}$ has minimum distance $d_2$.
Since $C_{2}\subseteq C_{1}\subseteq C_{1}^{\perp}\subseteq C_{2}^{\perp}$,
$C_{2}$ is self-orthogonal.
Hence the lemma has been proved.

\end{document}